\documentstyle[prl,aps,psfig,twocolumn,amssymb]{revtex}

\begin{document}

\twocolumn[\hsize\textwidth\columnwidth\hsize\csname @twocolumnfalse\endcsname
\bibliographystyle{plain}

\title{On the presence of mid-gap states in $\rm{CaV}_4{\rm O}_9$}
\vskip0.5truecm
\author {{\sc Marc Albrecht, Fr\'ed\'eric Mila} and {\sc Didier Poilblanc}}
\vskip0.5truecm
\address{Laboratoire de Physique Quantique\\
Universit\'e Paul Sabatier\\
31062 TOULOUSE \\
FRANCE}
\maketitle

\begin{abstract}
Using exact diagonalizations of finite clusters with up to 32 sites, we study
the $J_1-J_2$ model on the 1/5 depleted square lattice. 
Spin-spin correlation functions are consistent with plaquette order in the 
spin gap phase which exists for intermediate values
of $J_2/J_1$. Besides, we show that singlet
states will be present in the singlet-triplet gap if $J_2/J_1$ is not too
small ($J_2/J_1 \gtrsim 0.47$). We argue that this property should play a central
role in determining the exchange integrals in ${\rm CaV}_4{\rm O}_9$
\end{abstract}
\vskip2pc]
\narrowtext


The interest in 2D frustrated magnets has recently increased with the discovery 
of the first 2D spin 1/2 system exhibiting a spin gap, namely 
${\rm CaV}_4{\rm O}_9$~\cite{taniguchi}. 
This material consists of V$_4$O$_9$ 
planes separated by calcium layers\cite{bouloux},
each vanadium atom being at the centre of a VO$_5$ square-pyramid of oxygens.
Vanadium having an
oxidation number 4+ in this compound, there is a single electron in the
$d$-shell of each vanadium atom. These electrons behave 
as localized spins 1/2 coupled by some exchange interactions. The relative 
magnitudes of the various possible exchange mechanisms cannot be easily 
deduced from
quantum chemistry, and to get a better understanding of the magnetic
properties of this compound, a direct comparison of
physical properties of model Hamiltonians to experimental results
is necessary.

The minimal model to describe this compound, namely the 
1/5 depleted 
Heisenberg model 
with nearest neighbour interactions $J_1$, has been studied in great details.
Contrary to early results~\cite{katoh},
it is now clear that this model has N\'eel 
long range order and no spin gap. This was first proposed on the basis of exact
diagonalizations and Schwinger boson mean field calculations~\cite{albrecht}
and confirmed by Monte Carlo simulations~\cite{troyer} and 
other methods~\cite{ueda,miyazaki,starykh,gelfand}.

One way to go beyond this minimal model is to take into account the fact that
intra- and inter--plaquette exchange integrals might have different values, say
$J_1$ and $J_1'$, in
particular because
of the distortion of the lattice described in
Ref.\ \cite{bouloux}. However, it seems difficult to explain the magnitude
of the gap without assuming that the ratio $J_1'/J_1$ is unphysically small. 
 
\begin{figure}[ht]
\begin{center}
\mbox{\psfig{figure=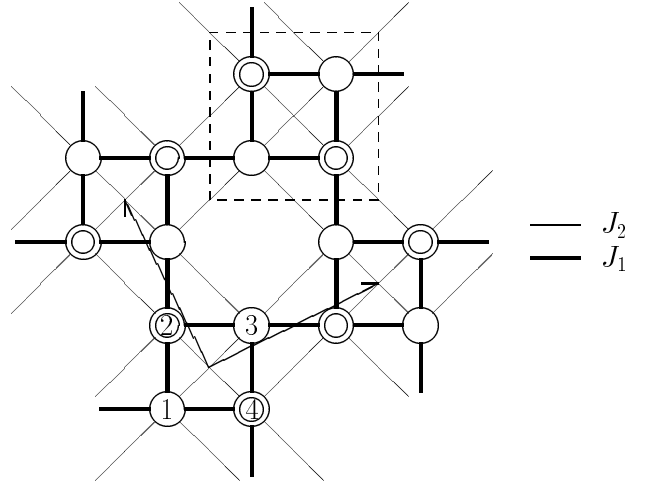,width=\columnwidth,angle=0}}
\end{center}
\caption{ Schematic structure of the V$_4$O$_9$
plane. The oxygen  have been omitted
for clarity.}
\label{lattice}
\end{figure}

In fact, in analogy with the non-depleted square lattice~\cite{heinz_review}, 
it has been suggested that a spin gap can be opened by frustration
e.g. through the inclusion of 
exchange integrals between next nearest 
neighbours~\cite{ueda,starykh,gelfand,white},
a very reasonable assumption as far as quantum chemistry is concerned.
This model corresponds to the Hamiltonian:
\begin{equation}  
H=  J_1  \sum_{<\!ij\!>} \vec{S_i}.\vec{S_j}  +
       J_2  \sum_{<\!\!<\!ij\!>\!\!>} \vec{S_i}.\vec{S_j} \ ,
\end{equation}
where $J_1$ ($J_2$) is the exchange integral between  
(next) nearest neighbours on the lattice shown in figure~\ref{lattice}.
Again, to account for the detailed properties of
${\rm CaV}_4{\rm O}_9$, it might prove necessary to allow for
different values of intra- and inter--plaquette exchange integrals, but these
differences can be neglected in a first approximation. 

A number of results have already been obtained on that 
model~\cite{ueda,starykh,gelfand,white}.
According to the 
``spin-wave''--like calculation of Starykh {\em et al.}~\cite{starykh},
there is a phase with a spin gap and plaquette order when 
$0.25 < J_2/J_1 < 0.8 $. 
The cluster expansion of Gelfand {\em et al.}~\cite{gelfand}
suggests that the model has a
plaquette--like ground state when $J_2/J_1=1/2$ with a large gap
$\Delta/J_1\approx 0.5$. This value of the gap was confirmed by
White~\cite{white} using a density matrix renormalization group (DMRG)
calculation, who also showed that the gap should open for a frustration ratio 
of $J_2/J_1\approx 0.05$, i.e. earlier than predicted by Starykh {\em et al}. 
Even if these studies are somehow convergent, a 
number of points remain to be definitely clarified.
First, the nature of the ground state in the
intermediate frustration domain has only been determined by perturbative
methods. In principle, they are valid when the perturbation
involves small parameters, which is not the case here. 
More importantly, the values 
of the 
coupling constants could not be deduced from experiments so far. 
In particular, the interpretation of the temperature dependence of the 
susceptibility is not
conclusive~\cite{gelfand}. Alternative ways of getting
information about the exchange integrals directly from
experimental results are clearly needed.

In order to address these issues, we have performed exact diagonalizations
of small clusters with 8, 16 and 32 spins with periodic boundary conditions.
To reduce numerical effort, we have taken advantage of all the 
symmetries of the
Hamiltonian: The translations (the basis vectors are shown in 
Fig.~\ref{lattice}), the point group $C_{4}$ (the centre of
rotation is at the middle of a plaquette) and the spin inversion. 
The elementary cell has four atoms labelled by $\alpha=1,2,3,4$ 
(see Fig.~\ref{lattice}). 
In the classical limit, the system will exhibit N\'eel order for $J_2/J_1<1/2$
and a collinear order with alternating rows (or columns) of up and down spins 
for $J_2/J_1>1/2$. 
A state with N\'eel (resp. collinear) long-range order will then
be defined by its wave vector ($\pi$,$\pi$) (resp. (0,$\pi$)) and by the 
orientation $\epsilon_{\alpha}=\pm 1$ of the 
spins of an elementary cell  
($\epsilon_1=-\epsilon_2=\epsilon_3=-\epsilon_4=1$ for N\'eel order,
$\epsilon_1=\epsilon_2=-\epsilon_3=-\epsilon_4=1$ for collinear order).

To test the presence of magnetic long-range order in the system,
we have calculated the ground state energy $E_N$, the singlet-triplet gap
$\Delta_N$, the magnetic susceptibility 
$\chi_N=1/(N\Delta_N)$, and the staggered magnetizations 
$M_N(\vec{Q})$ 
corresponding to N\'eel and collinear orders 
and defined by: 
\begin{equation}
 M^2_N(\vec{Q}) = \frac{1}{N(N+2)} \langle\Phi| \left( \sum_{i\alpha}
\epsilon_\alpha
e^{i\vec{Q}.\vec{r_i}}
\vec{S}_{i,\alpha}   \right)^2 |\Phi\rangle \  ,
\end{equation}
where $\vec{r_i}$ is the position of the centre of the plaquette $i$, and
$\epsilon_\alpha$ depends on the nature of the phase.
The normalisation of the staggered magnetizations
is chosen so that the order parameter is independent of the size
in a perfect N\'eel or collinear state. 
In an ordered phase, these various quantities should have 
the following finite size
scaling~\cite{neuberger}:

\begin{equation}  \frac{E_N}{N} = E_0 +  \frac{C_1}{N^{3/2}},  \hspace{1cm}
 \Delta_N =   \frac{(\chi_0)^{-1}}{N} \end{equation}
\begin{equation}  \chi_N=\chi_0 + \frac{C_2}{N^{1/2}}, \hspace{.5cm}
\ M_{N}(\vec{Q}) =M^{*}_{0}(\vec{Q}) + \frac{C_3(\vec{Q})}{N^{1/2}} \ .
\end{equation}

\begin{figure}[t]
\begin{center}
\mbox{\psfig{figure=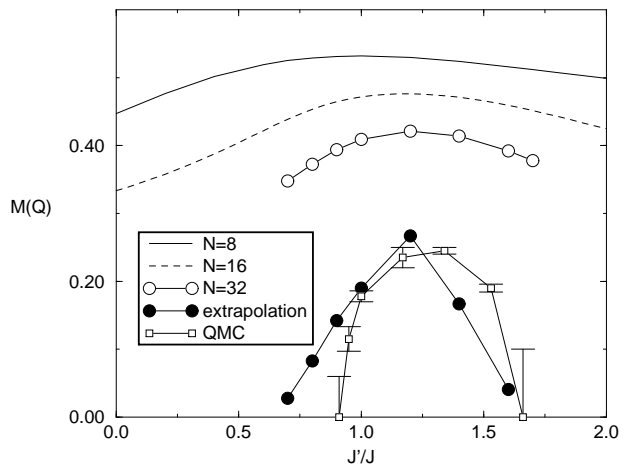,width=\columnwidth,angle=0}}
\end{center}
\caption{Staggered magnetisation of the non--frustrated model for N=8,16,32
and extrapolated values. Quantum Monte-Carlo 
results are from Troyer {\em et al.}\protect\cite{troyer}.}
\label{magnet1}
\end{figure}

Actually, for $J_2\ne 0$, these asymptotic laws are not very well 
satisfied for the available sizes
(8, 16 and 32 sites)~\cite{note1}, and the information we could extract 
on that problem is only qualitative. The ground-state energy has always 
a scaling reasonably well
described by Eq. (3), so it cannot really help deciding whether a state is 
ordered or not.
Finite size effects for the triplet gap and for the magnetic susceptibility,
for such sizes, become large when $J_2\ne 0$ so that no
reasonable scaling could be performed. 
However, the scaling of the staggered magnetization was satisfactory enough 
so that we were able to extract useful informations. 
Let us start with N\'eel order for the non-frustrated model
with intra- and inter--plaquette exchange integrals $J_1$ and $J_1'$ 
studied by two of us in a previous paper~\cite{albrecht}. 
We are now in a position to improve 
the finite-size analysis by considering the results for 32 spins. It turns out
that the scaling of Eq. (4) is not yet satisfied for such sizes. Meaningful
results can nevertheless be obtained by adding further corrections of order
$1/N$. 
They are
shown in figure~\ref{magnet1}.

\begin{figure}[ht]
\begin{center}
\mbox{\psfig{figure=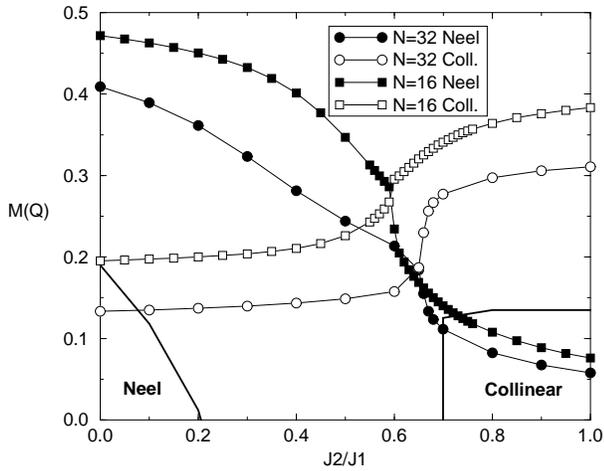,width=\columnwidth,angle=0}}
\end{center}
\caption{Staggered magnetisation and collinear order parameter
for the $J_1-J_2$ model for the N=16,32 sites clusters and 
extrapolated values (including also N=8 for the staggered case).}
\label{magnet2}
\end{figure}

 They are
in reasonable agreement with the Quantum Monte
Carlo results of Troyer {\em et al.}~\cite{troyer}. The slight discrepancy 
for small
$J'_1/J_1$ shows however that such a scaling should not be taken too seriously
at a quantitative level.
For the model of Eq. (1), the N\'eel staggered magnetization 
deduced from a similar finite size scaling analysis
is shown in figure~\ref{magnet2}.
It vanishes for
$J_2/J_1 \approx 0.2$. This is larger than the value 
$J_2/J_1 \simeq 0.05 $ deduced from
a recent density matrix renormalisation group (DMRG) calculation by
White~\cite{white}, which again suggests that large clusters have to be studied
to get quantitavite estimates.  
As far the collinear order is concerned, we first note that the 8 site cluster
cannot be used because the collinear order
is frustrated by periodic boundary conditions in that case. Using Eq. (4) with
16 and 32 sites, we found that 
there is a non--vanishing collinear order
parameter  for $J_2/J_1 \gtrsim 0.7$, in reasonable agreement with Starykh 
{\em et al.}~\cite{starykh}. This order parameter drops
abruptly, which suggests that the transition with the disordered state
is first order, as in the case of the non-depleted lattice~\cite{heinz_review}.
So, our results are consistent with the previous results that there is no
magnetic long-range order for intermediate values of $J_2/J_1$, although very
precise bounds cannot be deduced from a finite-size scaling of the results for
8, 16 and 32 sites.
 

Exact diagonalizations turn out to be very useful to study the nature
of the intermediate phase. The basic idea is that, for intermediate values of
$J_2/J_1$, the system will more or less behave as isolated plaquettes.
For $J_2/J_1=1/2$, this can be understood very simply for the following reason:
Let us write the Hamiltonian as $ H= H_{\rm P} + H_{\rm IP}$ where
$ H_{\rm P} $ describes independent plaquettes while 
$H_{\rm IP}$ describes the coupling between them, and let us denote by 
$|\Phi_0\rangle$ the ground
state of $H_{\rm P}$. Then, 
for $J_2/J_1=1/2$, $  \langle\Phi_0|H_{\rm IP}|\Phi_0 \rangle = 0 $, which
shows that the perturbation due to $H_{\rm IP}$ will have a very small
effect on $|\Phi_0\rangle$.

\begin{figure}[ht]
\begin{center}	
\mbox{\psfig{figure=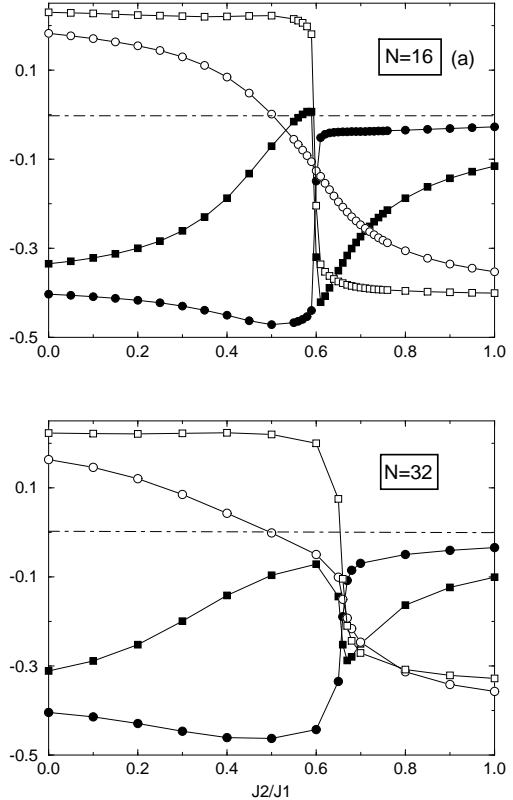,width=\columnwidth,angle=0}}
\end{center}
\caption{$\langle \vec{S_i}.\vec{S_j} \rangle$ on different links
for the N=16 (a) and  N=32 (b) clusters: P1 ({\large $\bullet$}), P2 ($\blacksquare$),
IP1 ({\large $\circ$}) and IP2 ($\square$).}
\label{orderparam}
\end{figure}

In order to check this assumption, we have computed 
$\langle\vec{S_i}.\vec{S_j}\rangle$ 
on different links of the lattice:
Along the side of a plaquette (P1), along the diagonal of a plaquette (P2), or
along inter--plaquette links (IP1 and IP2 for $J_1$ and $J_2$ respectively).
A signature of such a phase would be
a set of correlation functions  $\langle\vec{S_i}.\vec{S_j}\rangle$ close 
to their values in the pure plaquette phase~\cite{note2}.
Results for N=32 and N=16 are shown on
figure~\ref{orderparam}.
The general features of these correlation 
functions are
the same for both sizes. There is an abrupt variation of all the
spin correlations which occurs for $J_2/J_1 \approx 0.67$ for N=32 and 
$J_2/J_1 \approx 0.6$ for N=16. This difference in the ``critical''
frustration is the main finite size effect. These curves
give us information about the nature of the disordered phase: For 
$ 0.2 \lesssim J_2/J_1 \lesssim 0.7$, the correlations on all the links 
are close to their
values in the plaquette phase. The best agreement is for
$J_2/J_1=1/2$.
For this particular value of the frustration, the correlation functions on
the different links are comparable to their values in the pure plaquette state:
 -.4623 {\it vs} $-0.5$  for P1,
  .2201 {\it vs} $0.25$ for P2,  
  -.0963 {\it vs} 0 for IP1, and
 -.0009 {\it vs} 0 for IP2.
Hence, it seems clear that the ground-state is a plaquette resonating valence 
bond state. Note that there is no sign of the ground-state being degenerate, 
which allows one to eliminate the alternative broken symmetry state proposed by
Sachdev and Read\cite{sachdev}.

\begin{figure}[ht]
\begin{center}
\mbox{\psfig{figure=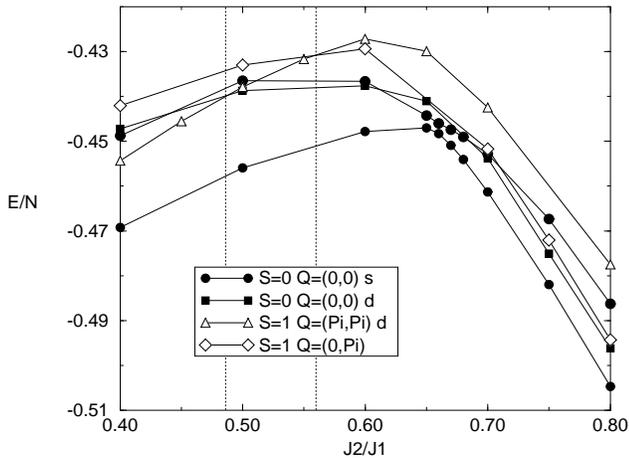,width=\columnwidth,angle=0}}
\end{center}
\caption{Lowest energy levels for N=32. Solid symbols are for spin 0
state, while open are for spin 1. Dashed lines indicate the position of the
change of the nature of the first excitations described in the text.}
\label{levels}
\end{figure}

We now discuss the low-lying excitations in the intermediate
frustration regime where the system has a plaquette ground state. 
For a single plaquette of four spins, the ground state is a
singlet of energy $-2J_1+J_2/2$ as long as $J_2<J_1$. 
However, the first excited state is a triplet only if
$J_2<J_1/2$. Beyond that value, there is a singlet state in the 
singlet-triplet 
gap. Let us see what remains of that picture when plaquettes are coupled.
The lowest energy levels versus $J_2/J_1$
for the $N=32$ spin cluster
are shown in figure~\ref{levels}. The other clusters have the same 
behaviour.
The ground state
is always a singlet totally symmetric under the symmetries of the Hamiltonian.
The nature of the first triplet excitation depends on the value of $J_2/J_1$.
When $J_2/J_1$ is small, the first triplet excitation has a 
wave vector $\vec{Q}=(\pi,\pi)$, while for $ J_2/J_1 \gtrsim 0.56 $ the
first triplet has a wave vector $\vec{Q}=(0,\pi)$.  
This change in the position of the lowest triplet in the Brillouin zone is
consistent with the prediction of Starykh {\em et al.}~\cite{starykh} and of
Gelfand {\em et al.}~\cite{gelfand} who both found such a change around
$J_2/J_1=1/2$.
A Schwinger boson mean field calculation~\cite{albrecht2} actually suggests 
that the
position of the maximum of the spin structure factor moves 
continously from $(\pi,\pi)$ to $(0,\pi)$.  
Now, more importantly, we found that, for intermediate values of $J_2/J_1$, 
there are singlet states in the
triplet gap when $0.49<J_2/J_1<0.7$. One of these excitations 
has the same wave vector as the ground state
but has a d--wave symmetry.
For the $N=16$ clusters, this energy level crosses
the ground state energy but this is probably an artifact of this 
cluster. 
The existence of a first excitation which is a singlet had not been reported
so far for this model.

Testing experimentally the presence of singlet states in the singlet-triplet
gap in ${\rm CaV}_4{\rm
O}_9$ 
should give useful information about the frustration
ratio for the following reasons. We first note that  
two related compounds have been
studied recently, suggesting that the value of the coupling constants could
be quite different from the commonly accepted value, {\em i.e.} $J_2/J_1
\approx 1/2$ and $J_1$ ranging from 100K to 200K. 
A parent compound of
${\rm CaV}_4{\rm O}_9$, ${\rm CaV}_3{\rm O}_7$,
which has a similar structure as far as the local
geometry in the VO plane is concerned, 
has been studied by Harashina {\em et
al.}~\cite{harashina}.  Using neutron diffraction, they have shown that it
has a collinear magnetic long range order. This is the signature of a large
frustration: 
A modified spin wave calculation of Kontani {\em et
al.}~\cite{kontani} suggests that this state is the ground state only if
$J_2/J_1$ is bigger than $0.7$. More recently, a quasi--one dimensional
vanadate, ${\rm NaV}_2{\rm O}_5$, has been studied~\cite{mila}.
This material consists of spin 1/2 chains corresponding to corner sharing
${\rm VO}_5$ square pyramids. So it involves only one coupling constant
which should correspond to $J_2$. A fit of the temperature dependence of the
susceptibility has been performed leading to $J_2 \approx 530K$.
The local
geometry being essentially the same in these compounds and in 
${\rm CaV}_4{\rm O}_9$,
the exchange integrals should be comparable. So these results suggest that 
$J_2$
should be of order 500 K, and that $J_2/J_1$ should be around 0.7. Note that
large values of $J_1$ and $J_2$  
are compatible with the reported value for the  
gap  ($107 K$) because the gap is much smaller
than $J_1$ close to the boundary to collinear order. 
Observing singlet states in the singlet--triplet gap would be a direct 
confirmation
of this picture.


We acknowledge very useful discussions with L. Levy and C. Lhuillier. 
The numerical calculations reported here were possible thanks to computing
time made available by IDRIS, Orsay (France).




\begin{references}
\bibitem{note1} In contrast to the non-depleted 
lattice\protect\cite{heinz_review}, the relevant scaling parameter 
is not, in fact, the number of sites but the number of unit cells, i.e. $N/4$.
\bibitem{note2} Unlike the non-depleted 
lattice\protect\cite{heinz_review}, no spontaneous (discrete) symmetry 
breaking is involved here.
\end{references}
\end{document}